\documentclass[conference]{IEEEtran}

\usepackage[british]{babel}
\usepackage[hyphens]{url}
\usepackage{graphicx}
\usepackage[noadjust]{cite}
\usepackage[pdftex,colorlinks=true,hyperfootnotes=false]{hyperref}
\IEEEoverridecommandlockouts

\begin{document}
%\bstctlcite{IEEEexample:BSTcontrol}

\title{``Share and Enjoy'':\\Publishing Useful and Usable Scientific Models}

\author{\IEEEauthorblockN{Tom Crick}
\IEEEauthorblockA{Department of Computing\\
Cardiff Metropolitan University\\
Cardiff, UK\\
Email: {\url{tcrick@cardiffmet.ac.uk}}}%
\thanks{The authors would like to acknowledge the use of the Sirius Cybernetics
  Corporation's motto for the title of this paper. }
\and
\IEEEauthorblockN{Benjamin A. Hall and Samin Ishtiaq and Kenji Takeda}
\IEEEauthorblockA{Microsoft Research\\
Cambridge, UK\\
Email: {\url{{benhall,samin.ishtiaq,kenji.takeda}@microsoft.com}}}}

\maketitle

\begin{abstract}
The reproduction and replication of reported scientific results is a
hot topic within the academic community. The retraction of numerous
studies from a wide range of disciplines, from climate science to
bioscience, has drawn the focus of many commentators, but there exists
a wider socio-cultural problem that pervades the scientific community.
Sharing code, data and models often requires extra effort; this is
currently seen as a significant overhead that may not be worth the
time investment.

Automated systems, which allow easy reproduction of results, offer the
potential to incentivise a culture change and drive the adoption of
new techniques to improve the efficiency of scientific exploration. In
this paper, we discuss the value of improved access and sharing of the
two key types of results arising from work done in the computational
sciences: models and algorithms. We propose the development of an
integrated cloud-based system underpinning computational science,
linking together software and data repositories, toolchains, workflows
and outputs, providing a seamless automated infrastructure for the
verification and validation of scientific models and in particular,
performance benchmarks.
\end{abstract}

% keywords
\begin{IEEEkeywords}
Reproducibility, Benchmarks, Models, Cloud Services, e-Infrastructure,
Computational Science, Open Science
\end{IEEEkeywords}

\IEEEpeerreviewmaketitle

\section{Introduction}

Two key types of results arise from work done in the computational
sciences: {\emph{models}} and {\emph{algorithms}}. Models represent an
abstraction of reality, and their behaviour is expected to be reliably
reproduced even if different algorithms are used. This validation of a
model's behaviour can be impacted by a number of factors relating to
the specific techniques used, but similar approaches are expected to
give broadly the same results.  In contrast, when new algorithms are
proposed to replace or supplement existing algorithms, they are
expected to verifiably replicate the results of other algorithms.

However, neither class of result exists in isolation: a new algorithm
is dependent on a set of models (or benchmarks) to demonstrate its new
capabilities. Equally, model development can both necessitate the
development of new algorithms and highlight the differences between
alternative approaches. Whilst algorithms and their implementations
have been highlighted as a potential barrier to
reproducibility~\cite{crick-et-al_wssspe2}, in this paper we discuss
the value of improved access and sharing of models in reducing
mistakes and in generating new scientific insights. We describe
efforts to reproduce computational models and algorithms, specifically
the multitude of issues relating to benchmarking of models and
algorithms.  We conclude with thoughts on where efforts should be
focused in both the short- and long-term to move to a world in which
computational reproducibility helps researchers achieve their goals,
rather than being perceived as an overhead.

%\section{Motivation}
We have seen a step-change in how science and engineering is
done. Experiments, simulations, models, benchmarks, even proofs cannot
be done without leveraging software and computation. A 2012 report by
the Royal Society stated that computational techniques have
``{\emph{moved on from assisting scientists in doing science, to
transforming both how science is done and what science is
done}}''~\cite{rssaaoe:2012}. Thus, the reproduction and replication
of reported scientific results is a widely discussed topic within the
scientific
community~\cite{barnes:2010,morin-et-al:2012,joppa-et-al:2013,goble:2014}.
Whilst the retraction of several studies has drawn the focus of many
commentators, automated systems, which allow easy reproduction of
results, offer the potential to improve the efficiency of scientific
exploration and drive the adoption of new techniques. Nevertheless,
this is a wider socio-cultural problem that pervades the scientific
community, with estimates that as much as 50\% of published studies,
even those in top-tier academic journals, cannot be repeated with the
same conclusions by an industrial
lab~\cite{osherovich:2011}. Furthermore, just publishing (linked)
scientific data is not enough to ensure the required
reusability~\cite{bechhofer-et-al:2013}.

Specific examples of the benefits of reproducible workflows to
researchers are now appearing in the scientific literature. For
example, new tools for membrane protein
simulation~\cite{Stansfeld,Hall2014} demonstrate how complex workflows
can be automated, preventing errors and differences arising from
manual execution, whilst making it faster to perform new
analyses. More complex tools, such as Copernicus, aim to automate more
generic molecular dynamics workflows~\cite{Pronk}. Alongside this,
recent work in executable biology~\cite{Cook} showed how a new class
of models, representing a defined property of biological networks,
defeated an existing algorithm for proving stability. As such, it was
the broader application of a new algorithm to additional models (or
benchmarks) which highlighted an unexplored but important phenomena
the algorithm could not address.

%Whilst experimental work requires detailed protocol descriptions,
%computer science and the computational science disciplines uniquely
%are able to share the raw outputs of their work as software and
%standard format datafiles. 
In each case, these tools take advantage of a fundamental advantage of
computer science and more broadly, computational science: the unique
ability to share the raw outputs of their work as software and
datafiles. However, despite this advantage, and alongside ongoing --
and significant -- changes to the traditional models of academic
dissemination and
publication~\cite{deroure:2010,stodden-et-al:2013,fursin+dubach:2014},
there remain cultural and technical barriers to both the sharing and
reimplementation of algorithms~\cite{crick-et-al_wssspe2}. These
include widely discussed topics, such as choices of programming
language and software licence, as well as unreported technical details
of the implementations and ensuring that research software developers
get credit for their work\footnote{e.g. Software Sustainability
Institute (\url{http://www.software.ac.uk}) and the UK Community of
Research Software Engineers (\url{http://www.rse.ac.uk})}. One
fundamental barrier to sharing remains: the overhead in time and
effort required to make models, algorithms and data reusable by a
third-party can be significant.

% SI: Don't use without permission. 
% \begin{figure}[!ht]
% \centering
% \includegraphics[width=\columnwidth]{phd031214s.png}
% \caption{The perils of reproducible research\newline [credit: \url{http://www.phdcomics.com/comics.php?f=1689}]}
% \label{fig:reprodres} 
% \end{figure}

However, even when this is considered, the testing of the benchmarks
themselves may be non-trivial. Benchmarks may be tailored to the
specific problems being addressed by the algorithm, and may not be
expected to cover all edge cases. Implementation details, such as
pseudo-random number generation, floating-point rounding behaviour and
order-of-execution, may affect benchmark results. Furthermore, for
high performance computing applications, repeating benchmark results
may not be possible by other groups without identical hardware
platforms and environments. Here we extend a previous
proposal~\cite{crick-et-al_wssspe2} by specifically discussing the
problems posed by models, considering the issues surrounding sharing
and analysing benchmark sets.

\section{The nature of models}

\subsection{Abstraction levels} 

A model describes reality at some level of abstraction. The more
detailed it is, the more ``faithful'' it often purports to be, but
also then the more special-purpose (and potentially less useful to
others). It is an important aspect of the modelling task as to what
level (abstract vs. concrete) to model at. But often it is implicit,
embodied but not embedded, in the model.

% Published models require basic information to allow the models to be
% explored. On the most basic level, this can include techniques to
% allow files to be parsed, but in this parsing errors may arise from
% details stored in the original implementations of the model analysis
% algorithms.  
An example of this comes from the treatment of floating-point
conversions in qualitative networks in systems
biology~\cite{Schaub2007}. Each vertex or variable in a network has an
algebraic target function which describes how the variable should
change at each step. The variables themselves are integers, and the
target function may return a float, which must be converted to an
integer for the update. This can be done in the target function
itself, but if the function returns a float, the specific
implementation dictates if this is a rounding, floor or ceiling
function. This implicit (to the model) but specific (from the
implementation) may change the results of the modelling. Another case
of showing how the implementation of qualitative networks may change
the model is the treatment of variable ranges within the model.
Whilst the formalism allows the variables in a model to have differing
ranges of integers, the mechanism of conversion is not specified in
the formalism. As such, this is another area where implementations can
dictate the precise behaviour of the model, and thus needs to be
explicitly annotated on the model.

A further example is the handling of pseudo-random number generation in
Avida~\cite{ofria+wilke:2004}, an open source scientific software
platform for conducting and analysing experiments with
self-replicating and evolving computer programs. In order to produce
consistent random number generation across platforms, it may be
necessary to code bespoke random number generators within the system,
which is not ideal for sharing and reproducibility.

%Eg., Int. Does it mean mathematical integers, C long, C long long, int32, uint64? 
%What about overflow behaviour? 
%e.g. What about random numbers?

\subsection{Benchmark repositories, curated}

% from Tom's section
% Benchmarks presented in scientific papers are essentially
% meaningless. If every paper has to be novel, then every benchmark,
% too, will be novel; there is no monotonic, historical truth in new,
% synthetically-crafted benchmarks. It is worse than that really: enough
% benchmarks are included to beat other tools. The comparisons are never
% fair (neither are other peoples' comparisons against your tool).  The
% benchmarks the tool describes are fashioned only for this instance:
% they might claim to be from the Windows device driver set, but the
% reality is that they are are stripped down versions of the
% originals. Stripped down so much as to be useless to anyone but the
% author vs. the referees. 

A benchmark is a set of models that have been put together for some
explicit purpose. Perhaps the directory structure of the benchmarks
indicates this purpose, perhaps assertions in the models indicate
this purpose; in short, the benchmarks need to be curated. If the
benchmark is public (allowing anyone to contribute), then the curation
is even more necessary to make the models reusable.

Once there is a set of tests, there is the issue of how independent
the tests are from each other.  The concept of ``composability'' is a
fundamental one in computer science.  Say we have two functions $f : A
\rightarrow B$ and $g : B \rightarrow C$. Their composition $f;g
: A \rightarrow C$ in some category with structure means that, if $f$
has property $P$ and $g$ has property $Q$, then $f;g$ has
property $P \oplus Q$, where $\oplus$ is some combinatorial operator
in the domain of discourse. What this abstract characterisation means
is that a program can be tested by testing its parts, whole system
testing can be done by unit testing.

But this is only the case if the system can be decomposed, and we know
that in many important areas, such as machine learning and
computational science, the models are often not decomposable.  We have
not been explicit about it -- people normally are not, especially for
models -- but we have been discussing algorithms and models assuming
that they are truly in-divisible objects. We have not required them to
be composable or decomposable. What can this mean practically? If we
have an algorithm $A$ that claims to run on model $M$ with result $R$,
then there is no reason to assume that a slight modification of $A$
will also have result $R$. Or that $A$ running on a $M * N$, for some
operator $*$, has a suitably extended $R$ result.

We thus need to be very careful when algorithms are running on models
automatically and asynchronously (or due to events beyond our
control), on a global scale, with an effect such as performance
results that matter to third-parties. Both algorithms and models will
need careful curating. Some good examples of such benchmarks are the
UCI Machine Learning
Repository\footnote{\url{http://archive.ics.uci.edu/ml/}}, Netflix
Prize benchmarks\footnote{\url{http://www.netflixprize.com/}}, SMT
Competition\footnote{\url{http://smtcomp.sourceforge.net/2014/}},
SV-COMP\footnote{\url{http://sv-comp.sosy-lab.org/2015/}}, Answer Set
Programming
Competition\footnote{\url{https://www.mat.unical.it/aspcomp2014/}},
and the Termination Problem
Database\footnote{\url{http://termination-portal.org/wiki/TPDB}}. Such
repositories would allow the tests to be taken and easily analysed by
any competitor tool. There has been some work towards developing this
connected infrastructure: for example, knowledge management systems to
preserve and share complete auto-tuning and machine learning setups
for optimisation, collecting all related artefacts and their software
and hardware dependencies besides just performance
data~\cite{fursin-et-al:2014}.

\section{Workflow of meta-models}

\subsection{Protocols as scripts}

Studying the behaviours of complex models is non-trivial. Whilst
concise methods sections of papers may give a representative minimal
working protocol (or workflow), missing details may present barriers
to its reproduction. This is exacerbated by the inclusion (or,
depending on the case, omission) of manual transformation steps which
may subtly change the model. Data format conversions may be
non-trivial and performed manually. These may involve ad hoc scripts,
which might not be part of any of the explicitly shared codebase. This
can be supported by open protocols, stored in electronic lab notebooks
during the process of model building. However even in this case,
assumed knowledge may prevent simple replication by a third party.
%Even better would be tools which aid the transformation of text-based
%protocols into real scripts.

One common approach to tackling complex protocols (or workflows) in
computational sciences is to automate the process by scripting the
laborious elements, such as in the Taverna Workflow Management
System~\cite{taverna:2013} for a range of disciplines from
heliophysics~\cite{leblanc-et-al:2013} to multi-disciplinary design
optimisation in engineering~\cite{crick-et-al:2009}.

A specific example for simulating molecular dynamics is
Sidekick~\cite{Hall2014Sidekick}: in its early steps, it builds an
initial model of a $\alpha$-helical peptide, performs an energy
minimisation of the peptide \emph{in vacuo}, solvates the peptide,
adds counter-ions, and runs a second energy minimisation. This is all
done without any user actions, and the subsequent replicates are
performed with different random seeds to collect accurate
statistics. Even in this ideal case however, variations may arise
between replicates. In testing on a hybrid AMD/Intel cluster, one of
the authors found that the solvation step added a variable number of
water molecules. In a molecular dynamics simulation, this is
sufficient to cause two simulations with otherwise identical starting
states to diverge over time. As such, the inherent properties of the
model and simulation should be taken into account in the overall
protocol design, and noted in any attempts to reproduce the
behaviour. Building, curating and sharing of scientific workflows can
provide consistently reusability, but this can require additional
effort on the scientist.  Workflows are often most effective where
scientific processes need to be repeatable many times, therefore
amortising the upfront cost of creating the workflow and its
components. This approach may not be appropriate for more exploratory
science, where the researcher tends to use a more interactive process
with their data and models. A successful example of community-building
in this space is the myExperiment project~\cite{myexperiment:2009},
which aims to make it easy to find, use and share scientific workflows
and other research objects.

In several disciplines, electronic lab notebooks have become the
norm. These tools, combined with open repositories such as
FigShare\footnote{\url{http://www.figshare.com/}} and
ZappyLab\footnote{\url{http://www.zappylab.com/}}, facilitate the
sharing of protocols. This may be needed for legal compliance
(e.g. drug trials), but has been successfully used in large research
consortia, for example the use of Accelrys Notebook (formerly Contur
ELN)\footnote{\url{http://accelrys.com/products/eln/contur/}} by the
structural genomics consortium in Oxford. Similarly, ZappyLab aims to
build a free, standardised protocol repository for the life
sciences. Within computational sciences, efforts to mine these
repositories could offer the potential to convert manual protocols and
work flows into prototype scripts, to aid reproducibility.

\subsection{Performance and scalability}

A key question in reproducing research in a computational context is
whether performance is a key issue. For models of the physical world,
such as computational fluid dynamics and molecular dynamics, it is the
resulting physics that is typically important to the end user, rather
than how fast it took to solve the computational problem. In
algorithms research, performance can be the key research result, and
therefore reproducing this is important. Another example is in high
performance computing where scalability of code
(e.g. GROMACS\footnote{\url{http://www.gromacs.org/}},
NAMD\footnote{\url{http://www.ks.uiuc.edu/Research/namd/}},
Desmond\footnote{\url{http://www.deshawresearch.com/resources_desmond.html}}). Here
the aim is to make simulations run more efficiently over large numbers
of cores/nodes. On-demand cloud resources such as Amazon Elastic
Compute Cloud, Google Compute Engine and Microsoft Azure offer
potentially attractive (and cost-effective) route to reproducing
computational experiments.

An important question is which performance metric to use. Wall clock
time is commonly used, but this does not allow for long-term
performance reproducibility as any such benchmarking is a snapshot in
time. This is true whether the underlying hardware the software is
running on is physical or virtual hardware. Some ``op count'' is a
more interesting measure. In many cases, the cost of hardware and
system artefacts are important but often overlooked, such as for
solvers in logic programming~\cite{brain+devos:2009}. Also, other
structural properties of the models the algorithms are running on, are
more interesting. In the field of systems biology, whether an
algorithm can prove properties like termination, stability,
interesting start conditions, etc, are useful measurements of whether
one algorithm is better than another. Recent initiatives such as the
Recomputation Manifesto~\cite{gent:2013}, explicitly overlooks
performance metrics, instead focusing on ensuring future
reproducibility\footnote{\url{http://www.recomputation.org/}}, with
runtime performance regarded as a secondary issue.

\section{Future outlook}

% Limitations presented by big data- will some models resist validation?

% Critical success factor is providing capabilities that help researchers 
% in their day-to-day work, rather than being an overhead incurred, thus 
% providing both immediate and long-term returns on time to the individual.

%\section{Continuous Integration for models}

The whole premise of this paper is that {\emph{algorithms}}
(implementations) and {\emph{models}} (benchmarks) are inextricably
linked. Algorithms are designed for certain types of models; models,
though created to mimic some physical reality, also serve to stress
the current known algorithms. An integrated autonomous cloud-based
service can make this link explicit.

In the software development world, no one would (should) commit to a
project without first running the smoke tests. You could be clever and
run the tests via the version control system's pre-commit hook. That
way you would never forget to run the tests. All of this can be done,
at scale, on the cloud now. Services such as
Jenkins\footnote{\url{http://jenkins-ci.org/}}, Visual Studio
Online\footnote{\url{http://www.visualstudio.com/en-us/products/what-is-visual-studio-online-vs.aspx}},
etc, schedule the tests to run as soon as you commit. We envisage
moving to a world in which benchmarks become available online, in the
same vein as open access of publications and research data. It seems a
small step to hook these continuous integration (CI) systems up to the
algorithm implementations that are written to run on these benchmarks.

Suppose you have come up with a better algorithm to deal with some of
these benchmarks. You write up the paper on the algorithm but, more
importantly, you also register the implementation of your algorithm at
this open service, as a possible algorithm to run on this benchmark
set. The benchmarks live in distributed git (or similar)
repositories. Some of the servers that house these repositories are CI
servers. Now, when you push a commit to your algorithm, or someone
else pushes a commit to theirs, or when someone else adds a new
benchmark, the service's CI system is triggered. It is also activated
with the addition of a new library, firmware upgrade, API change,
etc. All registered algorithms are run on all registered models, and
the results are published. The CI servers act as an authoritative
source, analogous to the Linux Kernel
Archives\footnote{\url{https://www.kernel.org/}}, of results for these
algorithms running on these benchmarks.

% other examples?
There are already several web services that nearly do all of this
things (for example, a repository for disseminating the computational
models associated with publications in the social and life
sciences~\cite{rollins-et-al:2014}), so a service that can integrate
most if not all of these features is possible. Such a service would
then allow algorithms and models to evolve together, and be
reproducible from the outset.

A system as described here has several up-front benefits: it links
papers more closely to their outputs, making external validation
easier and allows interested users to explore unaddressed sets of
models. Critically, it helps researchers to be more productive, rather
than being an overhead on their day-to-day work. In the same way that
tools such as GitHub make collaborating easier while simultaneously
allowing effortless sharing, we hope that we can design and build a
system that is similarly usable for sharing and testing benchmarks
online.

In summary, this proposed new infrastructure could have a profound
impact on the way that computational science is performed,
repositioning the role of models, algorithms and benchmarks and
accelerating the research cycle, perhaps truly enabling a ``fourth
paradigm'' of data intensive scientific
discovery~\cite{hey:2009}. Furthermore, it would effect the vital
cultural change by reducing overheads and improving the efficiency of
researchers.

% balance refs at the end
\IEEEtriggeratref{25}

\bibliographystyle{IEEEtran}
\bibliography{recomp2014}

\end{document}